\documentclass[english]{nature}
\usepackage[T1]{fontenc}
\usepackage[latin9]{inputenc}
\usepackage{amstext}
\usepackage{amssymb}
\usepackage{graphicx}
\usepackage{esint}

\makeatletter
\usepackage{graphicx}
\makeatletter
\let\saved@includegraphics\includegraphics
\AtBeginDocument{\let\includegraphics\saved@includegraphics}
\renewenvironment*{figure}{\@float{figure}}{\end@float}
\makeatother

\usepackage{siunitx}
\usepackage{braket}

\makeatother

\usepackage{babel}
\begin{document}

\title{Measuring the time a tunnelling atom spends in the barrier}

\author{Ram\'{o}n Ramos$^{1}$, David Spierings$^{1}$, Isabelle Racicot$^{1}$
\& Aephraim M. Steinberg$^{1,2}$}
\maketitle
\begin{affiliations}
\item Centre for Quantum Information and Quantum Control and Institute for
Optical Sciences, Department of Physics, University of Toronto, 60
St. George Street, Toronto, Ontario M5S 1A7, Canada.
\item Canadian Institute For Advanced Research, MaRS Centre, West Tower
661 University Ave., Toronto, Ontario M5G 1M1, Canada.
\end{affiliations}
\begin{abstract}
Tunnelling is one of the most paradigmatic and evocative phenomena
of quantum physics, underlying processes such as photosynthesis and
nuclear fusion, as well as devices ranging from SQUID magnetometers
to superconducting qubits for quantum computers. The question of how
\textit{long} a particle takes to tunnel, however, has remained controversial
since the first attempts to calculate it\cite{maccoll_note_1932,wigner_lower_1955},
which relied on the group delay. It is now well understood that this
delay (the arrival time of the transmitted wave packet peak at the
far side of the barrier) can be smaller than the barrier thickness
divided by the speed of light, without violating causality. There
have been a number of experiments confirming this\cite{ranfagni_delay-time_1991,Enders1992,steinberg_measurement_1993,Spielmann1994},
and even a recent one claiming that tunnelling may take no time at
all\cite{Sainadh2019}. There have also been efforts to identify another
timescale, which would better describe how long a given particle \textit{spends}
in the barrier region\cite{hauge_tunneling_1989,Landauer1994,Chiao1997}.
Here we present a direct measurement of such a time, studying Bose-condensed
$^{87}$Rb atoms tunnelling through a 1.3-\si{\micro}m thick optical
barrier. By localizing a pseudo-magnetic field inside the barrier,
we use the spin precession of the atoms as a clock to measure the
time it takes them to cross the classically forbidden region. We find
a traversal time of 0.62(7) ms and study its dependence on incident
energy. In addition to finally shedding light on the fundamental question
of the tunnelling time, this experiment lays the groundwork for addressing
deep foundational questions about history in quantum mechanics: for
instance, what can we learn about where a particle was at earlier
times by observing where it is now\cite{Steinberg1995,Steinberg1995a,Y.AharonovD.Z.Albert1988}?
\end{abstract}
While the earliest attempts to calculate the time a tunnelling particle
spends in the barrier region\footnote{not to be confused with the lifetime of a quasi-bound state}
addressed the propagation delay for a wave packet peak, work in the
1980s, particularly by B\"{u}ttiker and Landauer\cite{buttiker_traversal_1982,Buttiker1983},
shifted the discussion to an `interaction time', the time actually
spent by a particle in the barrier. This was motivated by the prediction\cite{hartman_tunneling_1962}
that in certain regimes, the wave packet delay (sometimes referred
to as the `phase time' or `Wigner time'\cite{wigner_lower_1955})
could appear to be superluminal, suggesting that it does not reflect
the duration of the tunnelling event. B\"{u}ttiker and Landauer provided
arguments in favour of a `semiclassical time' equal to $md/\hbar\kappa$,
where $m$ is the mass, $d$ is the barrier thickness, and $\kappa=\sqrt{2m(V_{0}-E)}/\hbar$
is the evanescent decay constant in the tunnelling regime ($E$ is
the incident energy and $V_{0}$ is the barrier height); it should
be noted that unlike the group delay, this grows linearly with $d$
and is thus generally not superluminal. As we shall discuss in depth
below, another way of defining the tunnelling duration is via the
effect a particle has on an auxiliary `clock' degree of freedom.

\begin{figure}
\begin{centering}
\includegraphics[width=1\columnwidth]{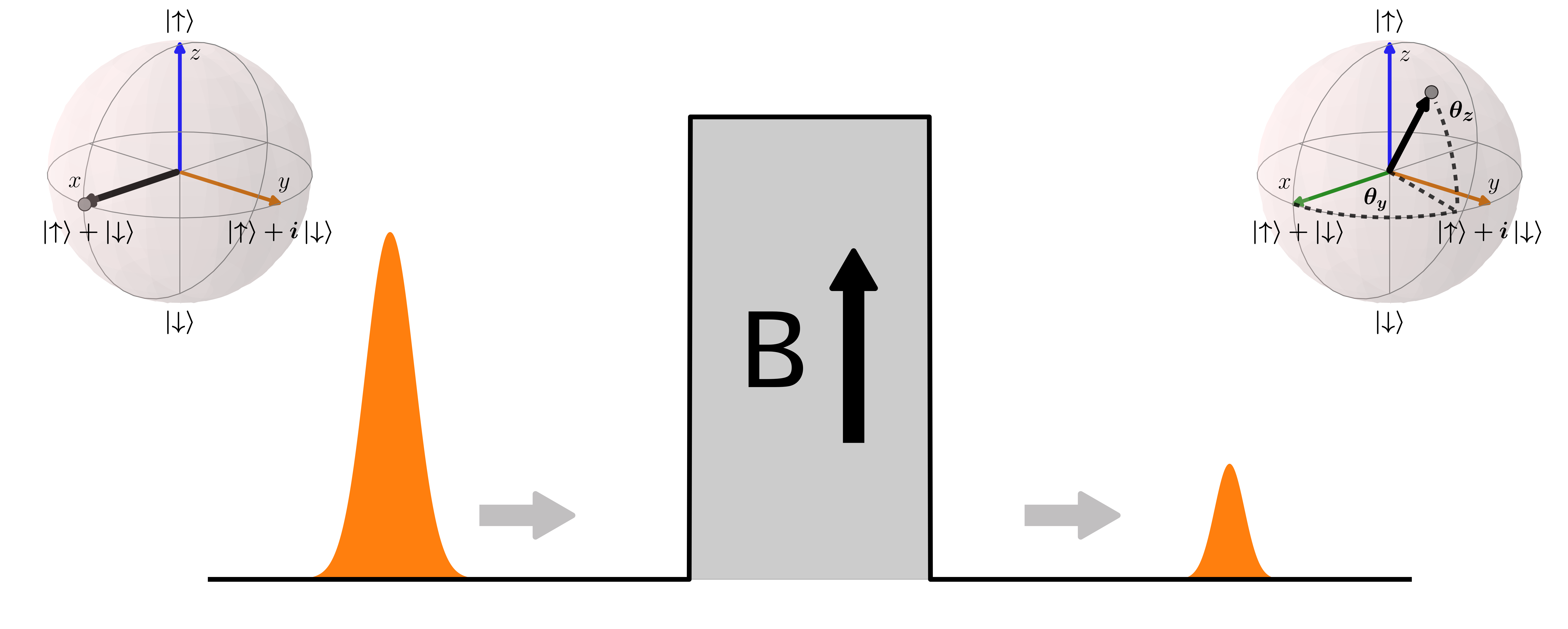}
\par\end{centering}
\caption{Larmor clock. A weak magnetic field $B$ pointing in the $z$-direction
is localized inside the potential barrier. A particle with spin-$1/2$
pointing initially along the $x$-direction impinges on the barrier,
and after transmission, the spin has precessed in the $xy$ plane
with a Larmor frequency $\omega_{L}$ and tilted towards the $z$-axis,
as depicted in the Bloch sphere. The $\ket{\uparrow}$ and $\ket{\downarrow}$
states are the eigenstates of the system in the presence of the magnetic
field along the $z$-axis. The Larmor times are then defined as $\tau_{y}=\theta_{y}/\omega_{L}$
and $\tau_{z}=\theta_{z}/\omega_{L}$.}
\label{Fig:1-Larmor}

\vspace{0.2in}
\end{figure}

A number of experiments have sought to measure tunnelling times. Several
optical experiments\cite{ranfagni_delay-time_1991,Enders1992,steinberg_measurement_1993,Spielmann1994}
have confirmed the superluminal nature of the group delay, and some
have attempted to probe other timescales\cite{deutsch_optical_1996,balcou_dual_1997}.
A pioneering experiment studying quantum tunnelling in a Josephson
junction\cite{Esteve1989} was the first attempt to apply the insights
of B\"{u}ttiker and Landauer to instead probe the duration of the tunnelling
event itself and offered qualitative agreement with the `semiclassical
time'. 

Recently, there has been an explosion of experimental interest in
the area due to the Keller group's development of the `attoclock'\cite{Eckle2008}.
These experiments\cite{eckle_attosecond_2008,Pfeiffer2013,Landsman2014,Zimmermann2016,Camus2017,Sainadh2019}
use strong-field ionisation in a circularly polarized field to determine
how much time elapses between the ionising field reaching its maximum
and an electron finally escaping. Intuitively, one might therefore
expect such times to be described by the group delay, although there
have been multiple theoretical approaches and no small amount of controversy.
In this setup, rather than impinging upon a barrier, the electron
escapes from a quasi-bound state\footnote{see the distinction made early on by Landauer\cite{Landauer1989}}.
Therefore, it is impossible to identify a time at which the event
`starts', as opposed to merely the moment at which the field reaches
its maximum. The problem is further complicated by atomic-physics
effects which generate additional delays unrelated to the tunnelling
event. The most recent experiment, attempting to properly account
for these, made the dramatic claim that tunnelling is essentially
instantaneous\cite{Sainadh2019}. There has also been an experiment
probing the time delay between two tunnel-coupled momentum components
of atoms oscillating in the wells of an optical lattice\cite{Fortun2016},
although it was unable to discriminate between the different theories. 

To provide an operational definition of the tunnelling time, it is
natural to devise a `clock' which ticks only while the particle is
present in the barrier region. The Larmor clock, originally introduced
by Baz'\cite{Baz1966} and Rybachenko\cite{Rybachenko1967}, is the
most famous example of such a thought experiment. If an ensemble of
polarized spin-$1/2$ particles impinges on a barrier, localizing
a magnetic field on the barrier region alone will cause the spins
to precess only while the particles are under the barrier (see Fig.
\ref{Fig:1-Larmor}). Considering incident particles polarized in
the $x$-direction and a magnetic field along $z$, one would expect
the spin to precess by an angle $\theta=\omega_{L}\tau$, where $\omega_{L}$
is the Larmor frequency and $\tau$ is the time spent in the barrier.
By working in the limit of a weak magnetic field ($\omega_{L}\rightarrow0$),
this time can be measured without significantly perturbing the tunnelling
particle. B\"{u}ttiker\cite{Buttiker1983} noted that even in this limit,
measurement back-action cannot be neglected, and results in preferential
transmission of atoms aligned with the magnetic field. This leads
to two spin rotation angles: a precession in the plane orthogonal
to the applied magnetic field, $\theta_{y}$, as well as an alignment
along the direction of the field, $\theta_{z}$. He defined times
associated with the spin projections: $\tau_{z}$, $\tau_{y}$ and
$\tau_{x}=\sqrt{\tau_{y}^{2}+\tau_{z}^{2}}$, the latter often known
as the `B\"{u}ttiker time'. It turns out that combinations of two such
quantities appear in other theoretical treatments as a single complex
number\cite{Pollak1984,Sokolovski1987}, but researchers were hesitant
to accept complex-valued times without a clear interpretation. Later,
Steinberg\cite{Steinberg1995,Steinberg1995a} associated $\tau_{y}$
and $\tau_{z}$ with the real and imaginary parts of the `weak value'\cite{Y.AharonovD.Z.Albert1988}
of a dwell-time operator, thereby providing them with distinct interpretations
as the inherent tunnelling time and the measurement back-action, respectively. 

We have built an experiment to implement the Larmor clock, making
use of the long de Broglie wavelengths achievable in atomic Bose-Einstein
condensates and the remarkable degree of control possible in such
systems, both for tailoring optical potentials and for manipulating
and measuring spins. The spatial resolution of the potentials is limited
only by the wavelength of the laser light used, and at our energy
scales \textendash{} which are on the order of $E/k_{b}\sim$1 nK
(10's of feV) \textendash{} the tunnelling probability is sizable,
while the millisecond-level timescales are convenient to probe experimentally.
A two-photon Raman transition driven by the barrier beam itself couples
the states of a pseudo spin-$1/2$ system, generating our Larmor clock:
the effective Larmor frequency $\omega_{L}$ is set by the two-photon
Rabi frequency $\Omega$ of the Raman transition. Using this scheme,
we are able to determine the barrier traversal time in our system
by measuring the final spin state of the transmitted atoms. 

\begin{figure}
\begin{centering}
\includegraphics[width=16cm]{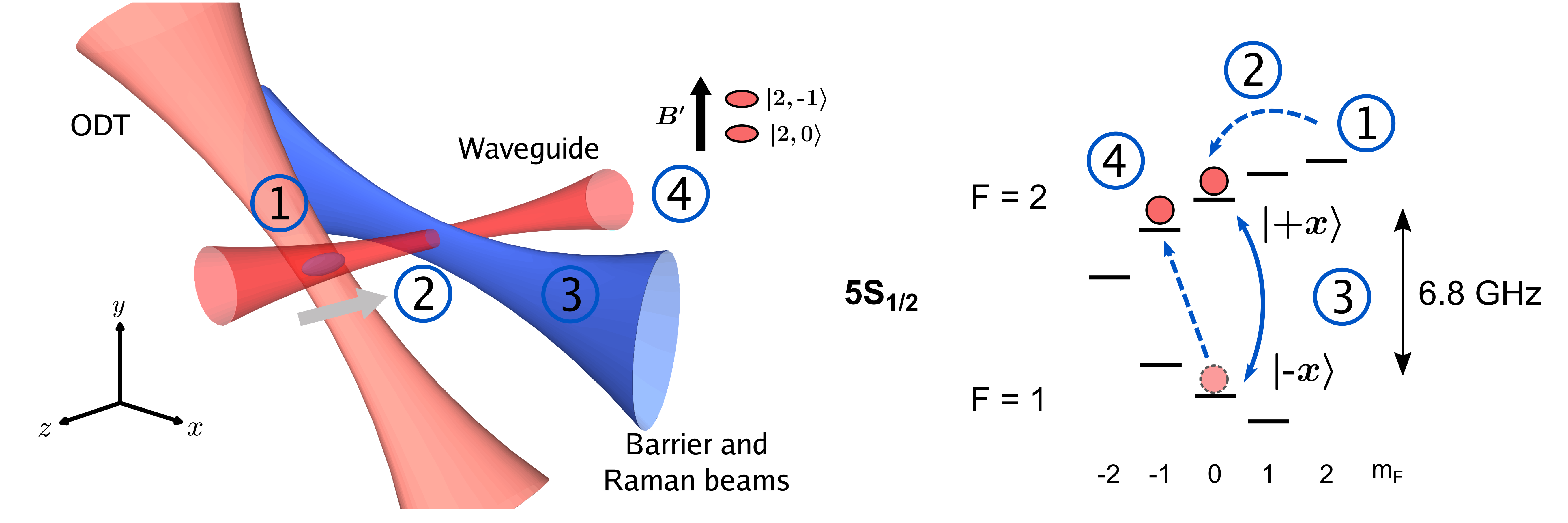}
\par\end{centering}
\caption{Experimental setup and sequence. We create a Bose-Einstein condensate
in the $|F=2,m_{F}=2\rangle$ state in a crossed dipole trap formed
by an optical waveguide and a perpendicular beam (ODT) intersecting
it. (1) After the effective temperature of the cloud is lowered, the
atoms are pushed, by a pulsed magnetic field gradient, along the waveguide
in the $\text{{-}}z$ direction towards a blue-detuned beam which
generates the potential barrier and the pair of Raman beams. (2) While
the atoms travel towards the barrier we use rapid adiabatic passage
to transfer them to the $\left|2,0\right\rangle $ $\left(\left|+x\right\rangle \right)$
state. (3) During the interaction with the barrier, the pair of Raman
beams couples the $\left|+x\right\rangle $ and $\left|-x\right\rangle $
states. (4) To perform the read-out sequence, the atoms which were
coupled to $\left|-x\right\rangle $ are transferred to the $\left|2,\text{{-}}1\right\rangle $
state, after which we perform a Stern-Gerlach measurement to separately
image both of the states.}
\label{Fig:Setup}

\vspace{0.2in}
\end{figure}

We prepare a degenerate gas of approximately 8000 $^{87}$Rb atoms
in the $5S_{1/2}$ $|F=2,m_{F}=2\rangle$ state ($F$ and $m_{F}$
denote the hyperfine and Zeeman quantum numbers respectively) in a
crossed dipole trap. One of the trap beams is turned off and the atoms
are left free to move longitudinally in a quasi one-dimensional waveguide.
We decrease the effective temperature of the atoms using matter-wave
lensing (see Methods), resulting in an rms velocity spread of 0.45(15)
mm/s, or an equivalent effective temperature of 2(1) nK ($\lambda_{\text{{dB}}}=$
4(1) \si{\micro}m). We then push the atoms towards a 1.3 \si{\micro}m
Gaussian barrier, formed by a focused blue-detuned beam (see Methods),
using a variable-duration magnetic gradient pulse to set the velocity.
We have previously observed tunnelling through this potential in two
different contexts: escape from a quasi-bound state\cite{Potnis2017,Zhao2017};
and in a single-collision geometry\cite{Ramos2018}, such as the one
studied here. While the atoms approach the potential barrier, rapid
adiabatic passage is used to transfer the atoms from their initial
spin state to the $|2,0\rangle$ state. An effective spin-$1/2$ is
encoded in the $\left|2,0\right\rangle $ and $\left|1,0\right\rangle $
hyperfine clock states denoted by $\left|+x\right\rangle =\frac{{|\uparrow\rangle+|\downarrow\rangle}}{\sqrt{{2}}}$
and $\left|-x\right\rangle =\frac{{|\uparrow\rangle-|\downarrow\rangle}}{\sqrt{{2}}}$
respectively (Fig. \ref{Fig:Setup}). The barrier light is phase-modulated
at the 6.8-GHz hyperfine frequency, thus creating a pair of Raman
beams that couple the $\left|+x\right\rangle $ and $\left|-x\right\rangle $
states. In the interaction picture, the pseudo-magnetic field generated
by the Raman beams points along the $z$-direction of the Bloch sphere.
After the collision with the barrier is complete, we perform a final
rapid adiabatic passage sweep from $\left|-x\right\rangle $ to $\left|2,\text{{-}}1\right\rangle $.
A subsequent Stern-Gerlach sequence separates the two spin states,
allowing us to determine the populations of $\left|\pm x\right\rangle $.

\begin{figure}
\begin{centering}
\includegraphics[width=16cm]{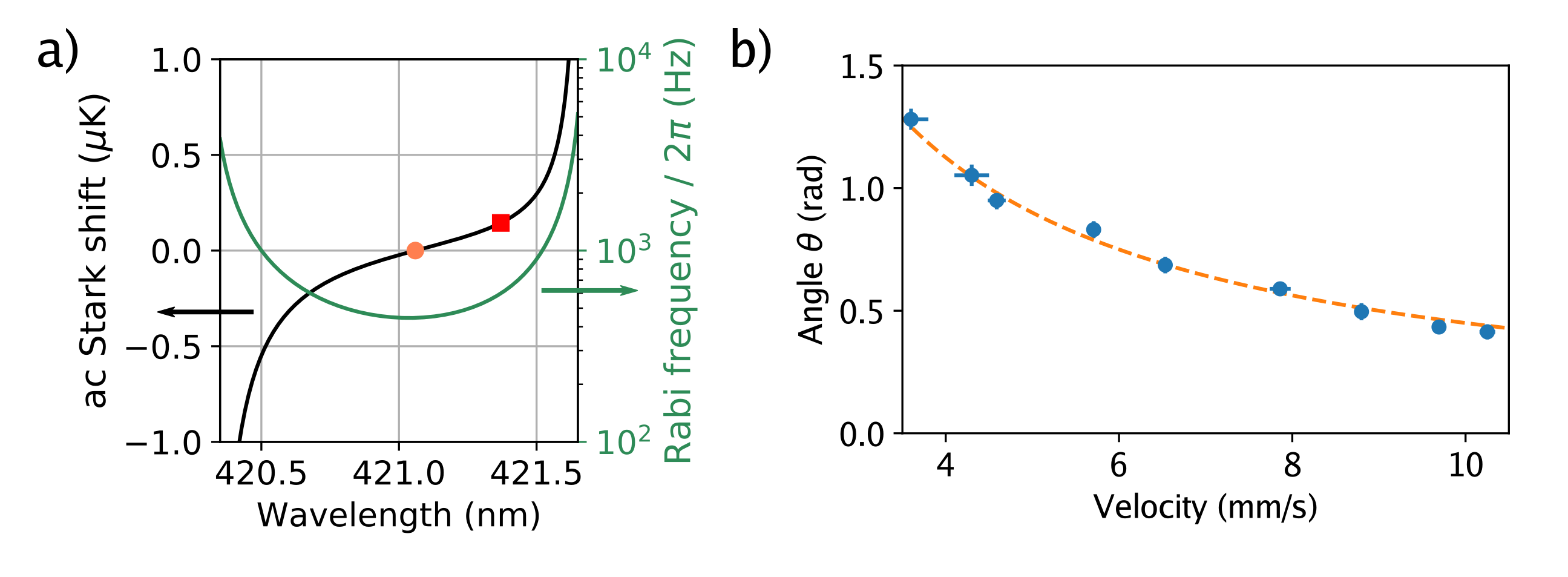}
\par\end{centering}
\caption{Larmor clock implementation. \textbf{a,} Scalar ac Stark shift (black
line) created by the tunable laser beam used for the optical potential.
The two points represent the wavelengths used in the experiment: the
orange dot is at the wavelength for which the barrier height vanishes
(tune-out wavelength) and where the free-space Larmor measurement
is performed, and the red square is at the wavelength used to create
a 135 nK potential barrier. The pair of Raman beams, with two-photon
Rabi frequency indicated by the green line, is created with the same
beam as the barrier. \textbf{b,} Measurement of the precession angle
at the tune-out wavelength. The dashed line is a one parameter fit
to the expression $\Omega d/v$, where $d$ is the effective barrier
width, $v$ is the incident velocity of the atoms and $\Omega$ is
the Rabi frequency, left as a free parameter. We find a Rabi frequency
of $\Omega=2\pi\times$ 440(10) Hz. The Rabi frequency $\Omega$ can
be controlled independently of the barrier height and was reduced
for the tunnelling data (see Methods).}
\label{Fig.: 2-Calibration}

\vspace{0.2in}
\end{figure}

We begin by testing the implementation of the Larmor clock in free
space. For this purpose, we remove the potential barrier but keep
the pseudo-magnetic field on by changing the barrier wavelength to
a nearby tune-out wavelength (see Fig. \ref{Fig.: 2-Calibration}
and Methods for details). The precession angle tells us how much time
the atoms spend in the region with the localized pseudo-magnetic field.
The results without the barrier (see Fig. \ref{Fig.: 2-Calibration}
b) show an expected $1/v$ dependence, where $v$ is the velocity
of the atoms. With our knowledge of the barrier width, we can extract
the Rabi frequency of the two-photon transition. 

After verifying the behaviour of the clock, we move to the case of
a repulsive barrier. In this case, after the interaction with the
barrier and the Raman beams, each atom will be either transmitted
or reflected, with its spin in a superposition of the $\left|+x\right\rangle $
and $\left|-x\right\rangle $ states (see Fig. 4a). At high energies,
the barrier has little effect on the atoms and the precession angle
scales inversely with velocity as before. From these points, we deduce
the Rabi frequency. We perform the tunnelling measurement with a 135(8)
nK barrier height, corresponding to 5.1 mm/s, and a Rabi frequency
of $\Omega=2\pi\times$ 220(40) Hz.

We investigate the two Larmor times by performing full spin-tomography
of the transmitted spin-$1/2$ particles. Rotations after the scattering
event allow us to measure the spin components along the $x$, $y$,
and $z$-axes of the Bloch sphere (see Fig. \ref{Fig:4-Data}b). From
the different projections, we find the traversal time $\tau_{y}$
and the time $\tau_{z}$ associated with the back-action of the measurement
(see Fig. \ref{Fig:4-Data}c). At the lowest incident velocity (4.1
mm/s), we observe a transmission probability of 3\%. Given the energy-dependence
of the transmission, we calculate that the transmitted atoms have
a velocity distribution with a peak at 4.8 mm/s, corresponding to
$\kappa d\simeq3$. About $3/4$ of this distribution corresponds
to energies below the barrier height. The measured traversal time
$\tau_{y}$ is 0.62(7) ms. 

The experimental data are in good agreement with one-dimensional two-component
time-dependent Schrödinger simulations with no free parameters (see
Fig. \ref{Fig:4-Data}c), while Gross-Pitaevskii simulations show
no significant modifications due to the presence of interactions.
Furthermore, the theoretical prediction given by the weak value formalism
describes our results well (see Methods). In contrast, the `semiclassical
time' disagrees with $\tau_{x}$ by more than three standard deviations
for the lowest velocities.

\begin{figure}
\begin{centering}
\includegraphics[width=0.65\columnwidth]{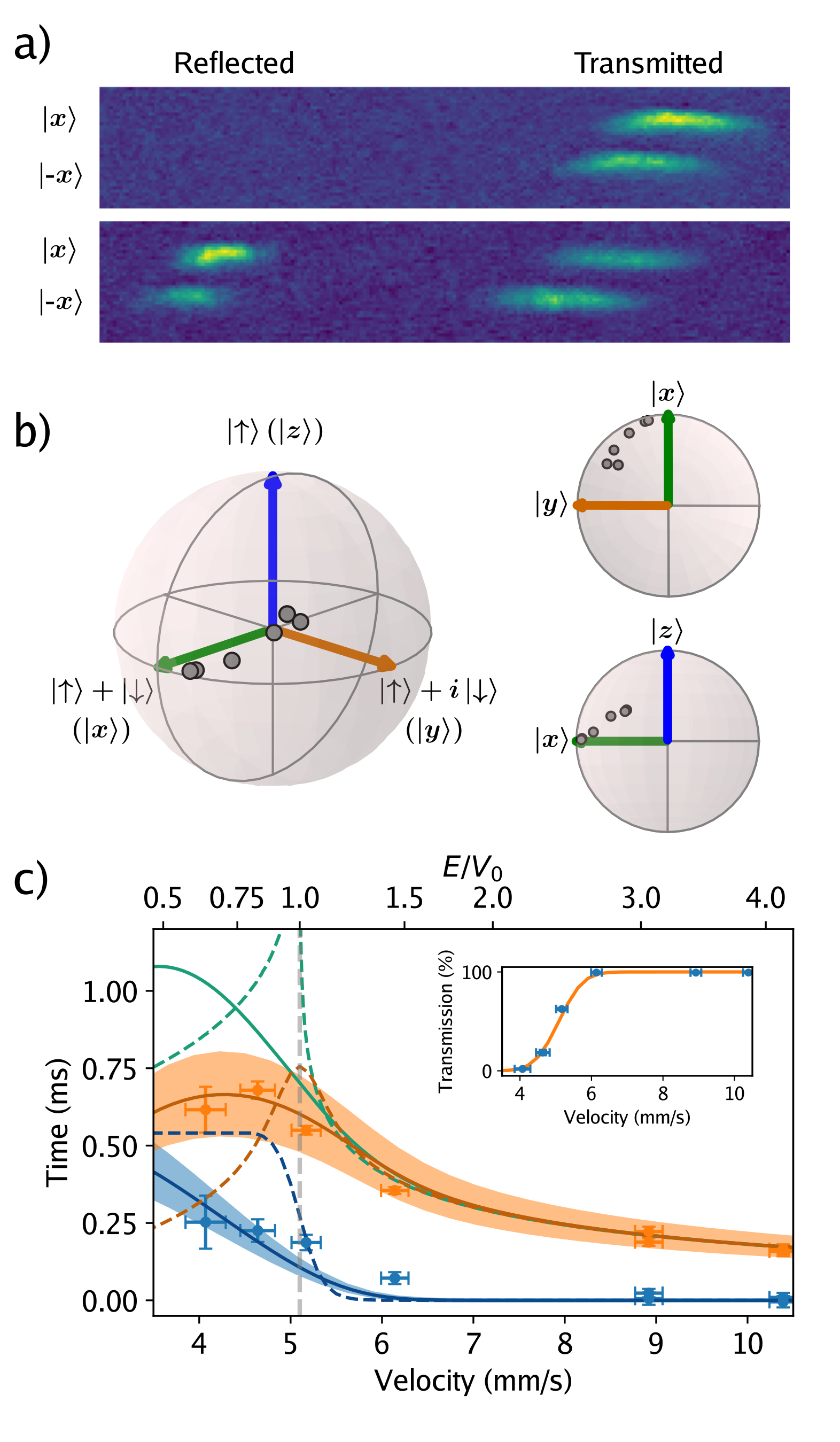}
\par\end{centering}
\vspace{0.2in}
\end{figure}

\begin{figure}
\caption{Traversal time of an atomic wavepacket through an optical potential.
\textbf{a,} Absorption images of the atomic densities after the interaction
with a 135 nK barrier and the Raman beams for incident energies of
400 nK (top) and 140 nK (bottom). The precession angles for the transmitted
atoms are obtained by doing full tomography on the spin-$1/2$ system.
\textbf{b,} Obtained spin projections for the different incident velocities.
The top right and bottom right figures show cuts along the $xy$ and
$xz$ planes. \textbf{c,} Experimental data of the Larmor times $\tau_{y}$
(orange dots) and $\tau_{z}$ (blue dots) as a function of incident
velocity. The dashed gray line corresponds to the velocity matching
the height of the barrier (5.1 mm/s). Light orange and blue regions
are 1D two-component time-dependent Schrödinger simulations, and the
bands represent the uncertainty on the measured Rabi frequency. The
dashed lines are monochromatic theory predictions for the Larmor times
$\tau_{y}$ (orange) and $\tau_{z}$ (blue), and the semiclassical
time (green). The corresponding solid lines are calculated by taking
into account the 0.45 mm/s velocity spread of the initial wavepacket.
\textbf{Inset}, Transmission probability data (blue dots) and two-component
time-dependent Schrödinger simulations using a 0.45 mm/s velocity
spread wavepacket (orange line).}
\label{Fig:4-Data}

\vspace{0.2in}
\end{figure}

This is, to our knowledge, the first direct measurement of the time
massive particles spend in a barrier region, implementing the Larmor-time
thought experiment of Baz', Rybachenko and B\"{u}ttiker. For a range of
incident velocities, we can observe both the time spent in the barrier
and the spin-rotation due to measurement back-action, clearly separating
the two effects. We see that as one heads deeper into the quantum
regime, the back-action grows in importance, and our results are consistent
with the prediction that the tunnelling time begins to decrease\cite{Buttiker1983}.
Our results are inconsistent with claims that tunnelling takes `zero
time'\cite{Sainadh2019,Eckle2008}. Beyond resolving the controversy
over how long a tunnelling particle spends in the barrier region,
the experimental approach pioneered here opens a new window on quantum
measurement and the broad question of how much one can infer about
the history of a quantum particle. In particular, it will enable measurements
of \textit{where} within a barrier transmitted and reflected particles
each spend their time\cite{Steinberg1995,Steinberg1995a}. As it is
predicted that these exhibit nonclassical behaviour, their study as
the system is made `more classical' by the introduction of dissipation
or atomic interactions promises to offer a new perspective on the
quantum-classical boundary. 
\begin{methods}

\end{methods}

\subsection{Experimental setup.}

We Bose-condense $8\times10^{3}$ $^{87}$Rb atoms in a 1064 nm crossed
dipole trap. The trap is composed of two beams: an elongated beam
which we refer to as the atom waveguide (with a waist of 15 \si{\micro}m
and a Rayleigh range of $z_{0}\approx$ 600 \si{\micro}m) and an
orthogonal beam intersecting the waveguide nearly 150 \si{\micro}m
away from its centre. The atoms start at the intersection of the two
beams. The atom waveguide has radial and longitudinal frequencies
of $\nu{}_{r}$ = 220 Hz and $\nu_{l}$ = 2.7 Hz. The barrier beam,
formed by a 421 nm laser focused to 1.3 \si{\micro}m (see below),
intersects the waveguide close to its centre. The longitudinal trap
frequency sets the minimum velocity at which the atoms meet the barrier,
since the waveguide curvature produces an acceleration of 5$\times10^{-2}$
m/s$^{2}$.

\subsection{Matter-wave lensing.}

We use delta kick cooling\cite{Chu1986,Ammann1997,Morinaga1999,Marechal1999,Myrskog2000},
also referred to as matter-wave lensing, to decrease the effective
temperature of the atoms from 15 nK to 2(1) nK, or a corresponding
rms velocity spread of 0.45(15) mm/s. At this temperature, the atoms
have a thermal de Broglie wavelength of $\lambda_{\text{{dB}}}=$
4(1) \si{\micro}m. The cooling procedure is as follows: the atoms
start in a crossed dipole trap, they are released from one of these
beams (ODT) and expand in the atom waveguide for 9 ms (to approximately
four times the initial cloud size), and then a quasi-harmonic trap
with a frequency of $\omega=2\pi\times$ 50(5) Hz created by the ODT
beam is flashed for 1.1 ms to collimate the atomic wavepacket. The
amount of expansion is limited by the radius of the beam (1/e$^{2}$
radius of 100 \si{\micro}m): this expansion time is kept short enough
that the atoms remain within the harmonic region of the Gaussian potential.
The condensate has a final longitudinal rms radius of 15 \si{\micro}m.
We have achieved temperatures as low as 0.9 nK using this technique
by reducing the initial atom number, but in order to work at higher
atom numbers, we perform this experiment at 2 nK.

\subsection{Potential barrier and Raman beams.}

The light for the barrier and the Raman beams is created by a homemade
421 nm external cavity diode laser (ECDL) in Littrow configuration.
We set the frequency of this beam in the vicinity of one of the rubidium
tune-out wavelengths (421.07 nm) located between the 5S$_{1/2}$ $\rightarrow$6P$_{1/2}$
(421.7 nm) and the 5S$_{1/2}$ $\rightarrow$6P$_{3/2}$ (420.3 nm)
transitions. To generate the pair of Raman beams, the beam passes
through a 6.8 GHz electro-optic phase modulator using a modulation
depth of $\beta\simeq$ 0.3 which creates a pair of sidebands in the
optical spectrum. About 5\% of the optical power goes to the sidebands.
The generated sidebands have opposite phases and act destructively
when used along with the carrier to drive Raman transitions. An etalon
with a FWHM bandwidth of 12 GHz is therefore used to remove one of
the sidebands. We can control the Rabi frequency without modifying
the power of the Raman beams, by adjusting the modulation depth $\beta$.
This is monitored by detecting the beat signal between the carrier
and the remaining sideband on a fast photodiode.

The barrier light is sent to the science chamber, where it is focused
to a 1/e$^{2}$ radius of $w_{z}$ = 1.3 \si{\micro}m along the $z$-direction,
while along the $y$-direction it is scanned with an acousto-optical
deflector to create a flat, 50 \si{\micro}m wide, time-averaged potential.
The Rayleigh range of the beam is 8 \si{\micro}m, larger than the
transverse radius of the condensate ($\sim$2.5 \si{\micro}m). We
can set the wavelength to 421.38 nm, where a power of approximately
0.5 mW generates a repulsive potential as large as $V_{0}/k_{b}$
\textasciitilde{}180 nK (given by the scalar ac Stark shift, as the
vector light shift vanishes for the clock states). A magnetic field
pointing along the propagation axis of the beams ($x$-direction)
sets the quantization axis. The beams are circularly polarized, to
drive $\sigma^{+}-\sigma^{+}$ Raman transitions. 

To calibrate the Rabi frequency, we study high-velocity incident atoms
traversing the barrier, which to a good approximation undergo free
propagation. The spin rotation due to the Raman beams is given by
$\theta=\int dt\,\Omega_{0}e^{-2(vt)^{2}/w_{z}^{2}}=\Omega_{0}d/v$,
where $v$ is the velocity of the atoms and $d=\sqrt{\pi/2}w_{z}$
is the effective barrier width. The Rabi frequency can also be calibrated
at the tune-out wavelength. Accounting for the difference in ac Stark
shift between the barrier and tune-out wavelengths, and frequency-dependent
transmission due to etaloning effects in our chamber windows, the
two techniques are in good agreement. However, due to the limited
tunability of the barrier ECDL, we utilize the first method to calibrate
the Rabi frequency.

\subsection{Larmor time calculations.}

We calculate the characteristic Larmor times using the weak measurement
formalism\cite{Y.AharonovD.Z.Albert1988}. The projection operator
onto the barrier region, $\Theta$, has eigenvalues 1 (for particles
in the barrier region) and 0 (for particles outside). A dwell time
operator $D\equiv\int_{-\infty}^{\infty}dt\Theta(t)$ provides a measure
of time spent in the barrier. However, its expectation value includes
contributions from both transmitted and reflected atoms. Steinberg\cite{Steinberg1995,Steinberg1995a}
showed that the weak value of this operator, $D_{w}=\tau_{y}+i\tau_{z}=\langle f|D|i\rangle/\langle f|i\rangle$
where $i$ refers to the initial state and $f$ to the final state,
can be understood as the conditional dwell time of a particle which
is prepared in the $|i\rangle$ state and postselected in the $|f\rangle$
state. In our experiment, the initial state corresponds to atoms incident
on the barrier from the left, while the final state corresponds to
transmitted atoms on the right side of the barrier. As originally
shown by B\"{u}ttiker\cite{Buttiker1983}, these Larmor times can also
be calculated as follows: $\tau_{y}=-\hbar\partial\phi/\partial V$
and $\tau_{z}=-\hbar\partial|T|/\partial V$, where $\phi$ and $|T|$
are the phase and the magnitude of the transmission amplitude, respectively.

Using the transfer matrix method, we solve for the conditional dwell
times at different incident energies numerically. In the experiment,
the Larmor probe is implemented by the pair of Raman beams; therefore,
in the calculations, we obtain the dwell times by integrating over
the Gaussian region of the barrier (the top-hat function $\Theta$
is replaced with a Gaussian weight function, representing the local
strength of the interaction with the clock). This integration region
extends beyond the turning points of the barrier for the range of
incident energies, and we calculate that about 40\% of the measured
time for the lowest incident energy comes from the time spent in the
classically forbidden region.

\subsection{Spin preparation, rotation and readout.}

After we accelerate the atoms using a 15 G/cm magnetic gradient pulse
for a variable time (0 - 0.9 ms), we prepare the spin state of the
atoms. While the atoms travel towards the barrier, we ramp up the
magnetic field to approximately 40 G. The high magnetic field causes
a difference in successive energy splittings of the $F=2$ manifold
of 210 kHz thanks to the quadratic Zeeman shift. This allows us to
transfer the atoms from $\left|2,2\right\rangle $ (in which they
Bose-condensed) to $\left|2,0\right\rangle $ using a 1 ms radio-frequency
rapid adiabatic passage. A multi-loop antenna provides the rf coupling
with a Rabi frequency of approximately $\Omega_{\text{{rf}}}\approx2\pi\times$9
kHz. The transfer efficiency is greater than 95\%, and the atoms left
behind can be identified and do not affect the subsequent dynamics
in the experiment. The total preparation sequence lasts for 9 ms.
Since the clock transition has a field dependence of 575 Hz/G$^{2}$,
we lower the magnetic field to 1 G prior to the interaction with the
Raman beams to reduce our sensitivity to magnetic noise. 

For the read-out sequence, we again increase the magnetic field to
40 G. Due to the quadratic Zeeman shift, the frequency difference
between the $\left|1,0\right\rangle $ to $\left|2,\text{{-}}1\right\rangle $
and the $\left|1,\text{{-}}1\right\rangle $ to $\left|2,0\right\rangle $
transitions is 110 kHz. We transfer the atoms through a rapid adiabatic
passage from $\left|1,0\right\rangle $ to $\left|2,\text{{-}}1\right\rangle $
in 1.5 ms. The transfer efficiency is 90\%. This sequence lasts for
8.5 ms. Subsequently, we lower the magnetic field to 1 G and perform
a 5.5 ms time-of-flight with a 40 G/cm magnetic gradient to implement
a Stern-Gerlach measurement and simultaneously image both of the final
states.

We can only directly measure the hyperfine states $\left|+x\right\rangle $
$\left(\left|2,0\right\rangle \right)$ and $\left|-x\right\rangle $
$\left(\left|1,0\right\rangle \right)$, thus obtaining $\langle Sx\rangle$.
To complete the tomography, after the interaction with the barrier
we apply a microwave pulse addressing the $\left|+x\right\rangle $
and the $\left|-x\right\rangle $ states to perform a $\pi/2$ rotation
along the $z$-axis to measure $\langle Sy\rangle$. Similarly, we
rotate the spin by a $\pi/2$ angle along the $y$-axis, through a
phase shift of $\pi/2$ radians in the driving field, to measure $\langle Sz\rangle$.
The rotations are done with a dipole antenna driving the atoms with
a Rabi frequency of $\Omega_{\text{{mw}}}=2\pi\times$2.4 kHz. To
identify and account for possible systematic errors such as imperfect
$\pi/2$ rotations or population transfer in the read-out sequence,
we also measure the -$x$, -$y$ and -$z$ projections. We periodically
calibrate the phase between the microwave source and the Raman beams
since drifts in the ECDL frequency and changes in the etalon due to
temperature fluctuations can occur. The phase calibration is performed
as follows: a spin rotation along the $z$-axis, for atoms with an
incident velocity well above the barrier height, is induced by the
Raman beams and it is followed by a microwave pulse tuned to provide
a $\pi/2$ rotation about a torque vector which lies in the $yz$
plane of the Bloch sphere, its angle on this plane depending on the
microwave source phase (relative to that of the Raman beams). We repeat
the procedure while scanning the phase of the microwave source and
determine the phase difference from the sinusoidal behaviour of the
measured hyperfine populations. For this calibration, we use a 120(7)
nK barrier and an incident velocity of 8.70(15) mm/s. The typical
phase drift between calibrations is 90 mrads and has been included
in the measurement uncertainties.

\bibliographystyle{naturemag}
\bibliography{nature-template,tunnelling_biblio_all_abbrv}
\begin{addendum}
\item We wish to acknowledge years of hard work by the students, postdocs,
and technologists who created this Bose-Einstein condensation apparatus
and helped turn it into the robust, precise machine which made the
present experiment possible: Ana Jofre, Mirco Siercke, Chris Ellenor,
Marcelo Martinelli, Rockson Chang, Shreyas Potnis, and Alan Stummer.
We thank Joseph Thywissen, Amar Vutha, Joseph McGowan, Kent Bonsma-Fisher
and Aharon Brodutch for valuable discussions. This work was supported
by NSERC and the Fetzer Franklin Fund of the John E. Fetzer Memorial
Trust. A.M.S. is a fellow of CIFAR. R.R. acknowledges support from
CONACYT.
\item [Author contributions]R.R., D.S. and I.R. performed the experiments.
A.M.S. supervised the work. All authors made contributions to the
work, discussed the results and contributed to the writing of the
manuscript. 
\item [Competing Interests] The authors declare no competing financial
interests. 
\end{addendum}

\end{document}